\documentclass{mem}
\usepackage{natbib}\usepackage{txfonts}\usepackage{balance}
\usepackage{graphicx}
\usepackage[a4paper,breaklinks,dvipdfm]{hyperref}
\idline{75}{282}
\begin{document}

\title{
Proper motions and brown dwarfs with the VVV survey
}


\author{
J.~C. \,Beamin\inst{1,2} 
\and V.~D. \, Ivanov\inst{2}
\and R.~ \, Kurtev \inst{3}
\and M.~ \, Gromadzki\inst{3}
\and K.~ \, Pe\~na Ramirez\inst{1}
\and D. \, Minniti\inst{1}
}

\institute{
Instituto de Astrof\'isica, Facultad de F\'isica, Pontificia Universidad 
Cat\'olica de Chile, Casilla 306, Santiago 22, Chile
\and
European Southern Observatory, Ave. Alonso de Cordoba 3107, Casilla 19001, Santiago, Chile
\and
Departamento de F\'isica y Astronom\'ia, Facultad de Ciencias, Universidad de Valpara\'iso, Ave. Gran Breta\~na 1111, Playa Ancha, Valpara\'iso, Chile
\email{jcbeamin@astro.puc.cl}
}
\authorrunning{Beamin}

\titlerunning{VVV Proper motion}

\abstract{
The Vista Variables in the V\'ia L\'actea survey (VVV) is a near-IR ESO public
survey devoted to study the Galactic bulge and southern inner disk covering
560 deg$^2$ on the sky. This multi-epoch and multi-wavelength survey has
helped to discover the first brown dwarfs towards the Galactic center, one 
of the most crowded areas in the sky, and several low mass companions to known
nearby stars. The multi-epoch information has allowed us to calculate precise
parallaxes, and put some constraints on the long-term variability of these objects.
We expect to discover above a hundred more brown dwarfs.

The VVV survey makes a great synergy with the Gaia mission, as both will observe for a
few years the same fields at different wavelengths, and as VVV is more sensitive to very red objects 
such as brown dwarfs,  VVV might provide unique candidates to follow up eventual astrometric microlensing
events thank to the exquisite astrometric precision of the Gaia mission.

\keywords{Stars: brown dwarfs--- 
methods: observational, 
techniques: imaging, spectroscopy, surveys, proper motions}
}
\maketitle{}

\section{Introduction}

Brown dwarfs (BD) are sub-stellar objects with very low surface 
temperatures (300$<$T$<$2200K) that are unable to sustain 
hydrogen fusion in their interiors. Low temperatures allow molecules
to play an important role in shaping the atmosphere spectrum of these objects, 
making colors at different wavelengths very distinctive. Most of previous 
successful searches were based on optical to mid IR color selections applied
to large area surveys uncloaking over two thousands of these objects \citep[e.g.][]{Kirkpatrick2011}.
As BDs are very cold and hence faint, most of the currently discovered BDs are
located relatively nearby. Then, an alternative and widespread way to discover these faint objects is
searching for high proper motion sources. Up to date several groups have worked on that with
 successful results \citep{Phanbao2008,Artigau2010, Lucas2010,Luhman2013,Luhman2014c} and the most recently 
catalogs of high proper motion objects based on WISE multiepoch data 
\citep{Luhman2014a,Luhman2014b, Kirkpatrick2014}. 
These efforts have unveiled the third and forth closest systems and the coldest brown dwarf discovered to date,
in addition to several other interesting ultra cool dwarfs.

The Vista Variables in the V\'ia L\'actea survey (VVV) is a near-IR multiepoch ESO public
survey covering 560 sq. degrees on the sky towards the Galactic bulge and southern disk.
\citep{Minniti2010,Hempel2014}
Although the main goal is to search for variable stars and study the 3-D structure of the 
Milky Way, it suits the requirements to measure proper motions and particularly 
the high spatial resolution allow us to spot and characterize new high proper motion sources.
Here we describe the methods, first results and perspectives of VVV survey search for high proper
motions sources and brown dwarfs and the connection with the observations carried out with the 
Gaia mission.

\section{VVV observations}

\begin{figure*}[ht]
\includegraphics{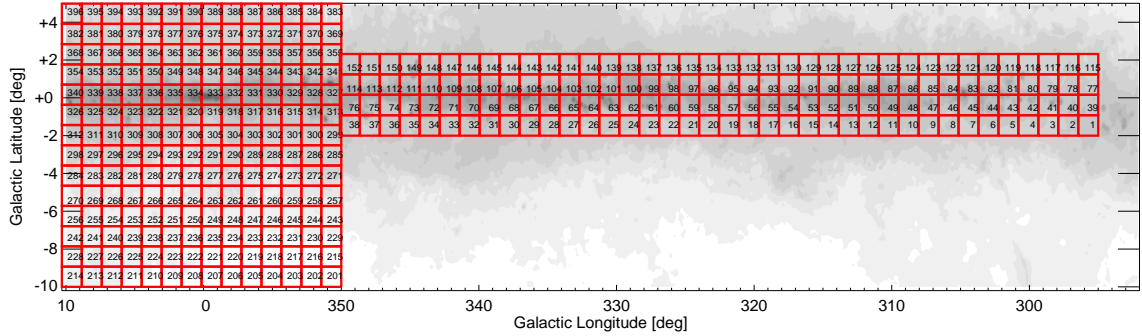}
\caption{
\footnotesize
VVV Survey area and tile numbering. The tile names start with
  ``b'' for bulge and ``d''  for disk tiles, followed by the numbering
  as  shown in the  figure. (Figure and captions from \citet{Saito2012}).}
\label{VVV_area}
\end{figure*}

The VVV survey is being carried out with VIRCAM at the VISTA telescope \citep{Emerson2006,Dalton2006} 
at cerro Paranal observatory. This instrument is made up of 16 detectors of 2048x2048 pixels, with a resolution of 
0.34\arcsec/pix. Each pointing, covering 0.6 deg$^2$, is called a `pawprint', and six overlapping 
pawprints are used to build one final image (Tile) covering twice an area of 
1.5 deg$^2$.
The image reduction, astrometric and photometric calibration and source catalog production are done by
 the Cambridge Astronomy Survey Unit\footnote{http://casu.ast.cam.ac.uk/surveys-projects/vista} \citep[CASU;][]{Irwin2004} 
The total sky area observed by VVV survey is 560 deg$^2$, and can be seen in Figure \ref{VVV_area}. The observations started at the end of 2009 and will finish
after 2016. The whole area has been observed in the $ZY$ $JHK_{S}$ bands and multiepoch observation in the $K_S$ band have
been obtained for every field with different cadence along the time. By the end of the survey the bulge area
should be observed around 100 times and the disk region slightly less times. 
These observations are well suited to calculate proper motion and derive parallaxes for nearby sources.

\section{Methods and current results} 
We started two searches of high proper motion sources (HPMs). The first one was based on visual inspection
of false color images using different $K_S$ epochs to: look for bright (probably saturated)\footnote{If the source is 
saturated the object might not be detected or might produce multiple detections or the accuracy of the centroid is 
compromised} sources that might been hidden before due to chance alignment in previous searches. We also 
search around previously known HPM stars to look for co-moving companions. We first found a new 'unusually blue' brown
dwarf at 17.5 $\pm$ 1.1 pc based on trigonometric parallax \citep{Beamin2013}. We also found 7 new companions to HPM stars
and obtain their spectral types. These new findings around HPM stars mean a 4\% incompleteness in the census of HPM  sources 
\citep{Ivanov2013}.

The second search was done using the catalogs alone. We cross matched four VVV $K_S$ epochs as evenly spaced in time 
as possible, then search for any object with motion larger than 0.07\arcsec /yr, we used four epochs to clean some 
spurious detections. We restricted our search to objects brighter than $K_S$=13.5 (more details in \citet{Gromadzki2013}).
Around 1500 HPM sources have been found and a final catalog describing these sources is being prepared (Kurtev et al. in prep.)

Also VVV data allow us to keep track of the long term photometric variability and the possibility to detect BD companions
and planets around M dwarfs and BD, this is a path we are now also exploring (Rojas-Ayala et al. submitted)

The final precision of the parallax measurements obtained with VVV data after the six years of observations, 
will be in the mili(sub-mili) arcsec regime, comparable to the results of other ground based astrometric measurements
\citep[among others]{Faherty2012,Marocco2013,Manjavacas2013,Smart2013}.

\begin{figure}[]
\resizebox{\hsize}{!}{\includegraphics[clip=true]{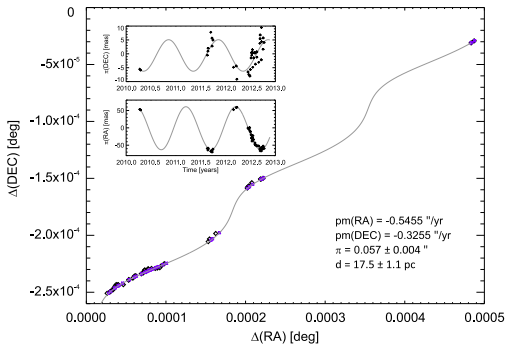}}
\caption{
\footnotesize
Parallax of \object{VVV BD001}, the distance was measured based on relative astrometry 
with a precision of a few mas ($\pi$=0.057\,$\pm$\,0.004 \arcsec), as more data become available, increasing the time
baseline and number of observations, the astrometric precision will improve to become close to 1 mas.
(Figure from \citet{Beamin2013}
}
\label{j}
\end{figure}

\section{VVV and Gaia synergy}
The VVV deeper near IR images and the new HPM objects found towards very crowded regions provide an excellent 
opportunity to find targets  for astrometric microlensing that could be followed up with Gaia. 
In some cases the superb astrometric precision of the Gaia mission will provide accurate independent estimates of the masses and
prove the existence of companions, planets or moons around these new objects and/or also presence of discs 
\citep{Dominik2000,Distefano2013,Sahu2014}.
The VVV also provide a unique complement to Gaia in the most crowded regions of the sky, as the number of sources that can be
 analyzed by the Gaia mission over those regions is ``fixed'' VVV will observe deeper and would complete most of other searches for
 proper motions and BD carried out by now, as well as the huge synergy in other fields of astrophysics such as galactic structure
 and bulge studies where VVV can go deeper and through most of the dust that would not allow Gaia to prove inner regions of the
 Milky Way. But on the other hand, Gaia will provide a deeper absolute astrometric reference frame with an exquisite precision
 for the proper  motions studies carried  out with the VVV observations.
 
\begin{acknowledgements}
Many thanks to the SOC/LOC and particularly to Ricky Smart for a very interesting and and enjoyable conference.
We acknowledge use data from the ESO Public Survey programme ID 179.B-2002 taken with the VISTA 
telescope, data products from  CASU. Funded by Project IC120009 ``Millennium Institute of Astrophysics (MAS)'' of 
Iniciativa Cient\'ifica Milenio del Ministerio de Econom\'ia, Fomento y Turismo de Chile.
 J.C.B., D.M., M.G., R.K., K.P.R., acknowledges support from: PhD Fellowship from CONICYT, Project FONDECYT
 No. 1130196, the GEMINI-CONICYT Fund allocated to project 32110014, FONDECYT through grants No 1130140, 3140351 respectively.
 This publication makes use of data products from, the Two Micron All Sky Survey, which is a joint project of the University 
 of Massachusetts and the Infrared Processing and Analysis Center/California Institute of Technology, funded by 
NASA and NSF.
\end{acknowledgements}

\bibliographystyle{aa}

\end{document}